\documentstyle[preprint,aps]{revtex}

\begin{document}

\title{ Dimensional versus cut-off renormalization  and the
nucleon-nucleon potential of the effective field theory}

\author {Sadhan K. Adhikari}
   \address{ 
  Instituto de F\'\i sica Te\'orica, 
Universidade Estadual Paulista, \\ 01405-900 S\~{a}o Paulo,  S\~{a}o Paulo, 
Brasil \\}

\date{\today }

\maketitle

\begin{abstract}

The  role of cut-off and dimensional regularizations  is discussed 
in the context of  obtaining  a renormalized nucleon-nucleon potential
from
the 
chiral Lagrangian formulation of the effective field theory  due to
Weinberg. 
Both types of renormalizations are
performed for the sum of  an attractive  delta
function and its
derivatives.   
The equivalence between the two forms of 
regularizations can be established with  the use of energy-dependent bare
couplings and 
the explicit forms of  these couplings are
determined.

{\bf PACS Numbers 21.30.-x,  03.65.Nk,  11.10.Gh}

\end{abstract}


\newpage

\section{Introduction\label{s1}}

The formulation and study of a nucleon-nucleon potential from a chiral
Lagrangian formulation of effective field theory due to
 Weinberg \cite{wei}
have become an important topic of investigation in nuclear physics
\cite{eft3}. The nucleon-nucleon potential derived from this
 effective field
theory contains usual finite-range potentials superposed on divergent
potentials containing delta function and its derivatives (gradients)
and can be written in momentum space as 
\cite{coh1,bir,eft2} 
\begin{eqnarray}\label{div}
V(p, q)&=& V_f(p, q)+
\lambda_1+\lambda_2(p^2+q^2)
+\lambda_3 p^2 q^2\nonumber \\
&+&\lambda_4(p^4+q^4)
+ \ldots,
\end{eqnarray}
where $V_f(p, q)$ represents usual finite-range parts of the potential. 
The constant term $\lambda_1$ is the $\delta$ function. 
The configuration-space derivatives of the $\delta$ function appear as 
powers of momenta in momentum space. These derivatives are the latter
terms of (\ref{div}) with coefficient $\lambda_i$, $i>1$.
Apart from the finite-range potential
$V_f(p, q)$, all other terms of potential (\ref{div}) are of zero-range
and possess ultraviolet divergences in momentum space. There is no
convergent calculational scheme with these divergent terms in the
Schr\"odinger framework. Meaningful solution is only obtained after
regularization and renormalization of the dynamical equations
\cite{coh1,bir,eft2,ad,adx,lat,sad}. 

Lately,
there have been a series of studies directed towards a successful
regularization and renormalization of potential (\ref{div}) and similar
divergent potentials
\cite{coh1,bir,eft2,ad,adx,lat,sad,jac}. 
 Three schemes have been used for the purpose: cut-off
regularization 
\cite{coh1,bir,eft2,ad,adx,sad,jac},  
dimensional regularization \cite{coh1,bir,eft2,sad},
and discretization on the lattice \cite{lat}.  
For the simplest
$\delta$-function potential all three approaches lead to identical result.
Although, the renormalization of the $\delta$ function term is
straightforward and completely under control, inconsistencies appear as 
the derivatives of the $\delta$ funtion are included in the
renormalization
scheme. 

Great deal of effort has been spent on the renormalization of the 
following part of potential (\ref{div}) \cite{coh1,bir,eft2,sad}
\begin{equation}\label{3}
V(p, q)=
\lambda_1+\lambda_2(p^2+q^2),
\end{equation}
where we include the sum of the $\delta$ function and its first
derivatives. Both dimensional and cut-off regularizations have been
applied for this purpose. However, no satisfactory and
consistent renormalized result for the solution has been obtained with 
potential (\ref{3}). In some approach the large nucleon-nucleon scattering 
length sets a new scale at which a perturbative approach breaks down. This
scale corresponds to a very low energy.

The dimensional regularization of potential (\ref{3}) is immediate and
simple. But this result could not simulate the general low-energy behavior
of the two-body scattering amplitude. The cut-off regularization of 
potential (\ref{3}) involves messy algebra and it is difficult to
define a bare coupling which will lead to a satisfactory renormalized
result consistent with that obtained by dimensional regularization. An
excellent account of this controversy is given in Refs. \cite{coh1}.

Two approaches have emerged to avoid these related  problems. Kaplan,
Savage, and Wise \cite{kap} suggested the so-called power divergent
subtraction
scheme based on a effective Lagrangian with nucleons and pions including 
contact interaction. The divergent integrals are then treated via
dimensional regularization with an unusual subtraction scheme.
Consequently, they are able to obtain a convergent perturbative scheme for
large scattering length. Subsequently, Cohen and Hansen \cite{coh}
achieved
the same
result via a conventional cut-off regularization procedure in
configuration space. An interesting discussion on this has appeared in
Ref. \cite{new}.

Ultarviolet divergences also appear in perturbative quantum field
theory and are usually treated by renormalization techniques.
There are several variants of renormalization which employ different types of
regularizations,  such as,  the cut-off, and dimensional regularizations.
Usually,  both regularization schemes  lead
to the same renormalized result at low energies. The closely related
technique   of discretization on the lattice in such field theoretic problems
also should lead to equivalent results.   In view of this the discrepancy
and inconsistency found in the renormalization of potential
(\ref{3}) are quite alarming. 

The purpose of the present work is to perform a satisfactory
renormalization  in momentum space which is devoid of the problem
and inconsistency mentioned above.
The ultraviolet divergences encountered in the solution of the scattering
problem with potential (\ref{3}) are energy dependent.
In the usual text-book renormalization problems these divergences are
energy independent and one uses an energy-independent bare coupling for 
renormalization. 
Recently, we have suggested \cite{ad,sad} that for a satisfactory and
consistent 
renormalization  of potentials with energy-dependent ultraviolet
divergences it is advantageous to use energy-dependent bare coupling.
By exploiting the  flexibility  obtained with the use of energy-dependent
bare coupling, it is possible to perform  a general and consistent  
renormalization of potential (\ref{3}). 
In this work we follow this procedure explicitly, suggest explicit forms
for the bare couplings,  and perform the
renormalization of the potential (\ref{3}) by employing 
energy-dependent bare coupling. The present procedure leads to a
general scattering solution for a short-range potential and establishes
consistency between cut-off and dimensional regularizations.

In Sec. II we renormalize the $K$ matrix with potential (\ref{3}) using
dimensional and cut-off regularizations and in Sec. III  we present a
summary of the present investigation.

\section{Regularization and Renormalization}

The partial-wave Lippmann-Schwinger equation for the $K$ matrix
$K(p, q, k^2)$,  at center of mass  energy $k^2$,  is given,  in three
dimensions,  by
\begin{eqnarray}
K(p, k, k^2) &=& V(p, k)\nonumber \\
&+& {\cal P}
\int q^{2} dq V(p, q)
G(q; k^2)K(q, k, k^2),\label{2}
 \end{eqnarray} 
with the free Green function $G(q; k^2)=(k^2-q^2)^{-1}, $ in units
$\hbar=2m=1$,  where $m$ is the reduced mass;  ${\cal P}$ in  Eq. (\ref{2})
denotes principal value prescription for the 
integral and the momentum-space integration limits are from 0 to $\infty$. 
The (on-shell) scattering amplitude 
$t_{L}(k)$ is defined  by 
\begin{eqnarray}
\frac{1}{t(k)}=\frac{1}{K(k^2)}+i\frac{\pi}{2}k,\label{33} 
\end{eqnarray}
where $K(k^2)\equiv K(k, k, k^2)= -(2/\pi)(\tan \delta/k)$ with 
$\delta$ the phase shift.
All scattering observables can be calculated using ${t({k})}.$
The
condition of unitarity is given by
\begin{eqnarray}
\Im t(k) = -\frac{\pi}{2}k |t(k)|^2,\nonumber
\end{eqnarray}
where  $\Im$ denotes the imaginary part. Here  we employ a $K$-matrix
description of scattering. Then the renormalization algebra will involve only
real quantities and we do not have to worry about unitarity which can be
imposed later via Eq. (\ref{33}). This is the simplest procedure to follow,  as
all  renormalization schemes preserve unitarity.

Now we consider the renormalization of potential (\ref{3}),
which is the sum of
a $\delta$ function and its second derivatives in configuration space. 
After a straightforward calculation,  
 the on-shell $K$ matrix for this potential is given by \cite{sad}
\begin{eqnarray}
 &K&( k^2) \nonumber \\
&=& \frac{{2k^2}/{\lambda_2}+{\lambda_1}/{\lambda_2^2}
+I_2(k^2)-2k^2 I_1(k^2)+k^4 I_0(k^2)}
{{1}/{\lambda_ 2^2}-{2I_1(k^ 2)}/{\lambda_2}+I_1^2(k^2)
-{\lambda_1 I_0(k^2)}/{\lambda_2^2}-I_0(k^2)I_2(k^2)},\label{a10}
\nonumber \\ \end{eqnarray}
where $I_L(k^2)$'s are the following regularized integrals
\begin{eqnarray}\label{20}
I_L(k^2)={\cal P} \int q^2 dq q^{2L} G_R(q;k^2),
\end{eqnarray}
where $G_R(q;k^2)$ is some regularized Green function. One can employ
cut-off and dimensionally regularized Green functions to calculate the
regularized integral (\ref{20}).  If one employs the following regularized
Green function with a sharp cut off $\Lambda$
\begin{eqnarray}
G_R(q,\Lambda;k^2)=(k^2-q^2)^{-1}\Theta(\Lambda -q),\nonumber
\end{eqnarray}
with $\Theta(x) =0$ for $x<0$ and = 1 for $x>0$,
the integral in Eq. (\ref{20})  can be analytically evaluated and one has
the following renormalized result
\begin{eqnarray}
I_{L}(k^2,\Lambda) &=& -\left[\sum_{j=0}^L\frac
{{k^{2(L-j)}}{\Lambda^
{2j+1}}}{2j+1} +\frac{k^{2L+1}}{2}\ln \biggr|\frac{\Lambda - k}{\Lambda
+k}
\biggr| \right],\nonumber \\
&=& -\left[ \Lambda + \frac{k}{2}\ln \left|\frac{\Lambda - k}{\Lambda
+k}\right|
\right],
 L=0,\nonumber \\
&=& -\left[ \frac{\Lambda^3}{3} + k^2\Lambda  +\frac{k^3}{2}\ln
\left|\frac{\Lambda - k}{\Lambda +k}\right| \right],
   L=1,\nonumber \\
&=& -\left[ \frac{\Lambda^5}{5}+\frac{k^2\Lambda^3}{3}+k^4\Lambda +
\frac{k^5}{2}\ln \biggr|\frac{\Lambda - k}{\Lambda +k} \biggr|
\right], L=2.\nonumber 
\end{eqnarray}
These integrals can also be treated by dimensional regularization. The
dimensionally regularized results for these integrals in three
dimensions are \cite{sad}
\begin{equation}\label{40}
I_{L}(k^2)=-\frac{1}{2} \Gamma\left( \frac{2L+3}{2} \right)
\Gamma\left(\frac{-2L-1}{2} \right) \Re [(-k^2)^{(2L+1)/2}]
\end{equation}
where $\Re$ is the real part.  As this result involves the real part of an
imaginary quantity it is identically zero in three dimensions.

First we perform cut-off regularization in Eq. (\ref{a10}).
This expression   has to be  calculated by introducing a cut off 
$\Lambda$ in the Green's function. However, in the end the limit 
$\Lambda \to \infty$ has to be taken. If the couplings 
$\lambda_1$ and $\lambda_2$ are taken to be constants, the
ultraviolet divergence appears in Eq. (\ref{a10})  as this limit is taken.
For a proper renormalized result to be obtained 
 the parameters  $\lambda_1$ and $\lambda_2$ of Eq. (\ref{a10})
are to be interpreted as
cut-off ($\Lambda$) and energy-dependent bare couplings. This $\Lambda$
dependence of $\lambda_1$ and $\lambda_2$ cancels the divergent
parts of the result in (\ref{a10}) when the $\Lambda \to \infty$ limit is
taken and one obtains a finite $K$ matrix. It is easy to write the
explicit form of the bare couplings, which are
\begin{eqnarray}\label{55}
\frac{\lambda_1}{\lambda_2^2}&=& -I_2(k^2,\Lambda)
+\Lambda_0(k^2)k^4,\\
\label{56}
\frac{1}{\lambda_2}&=& I_1(k^2,\Lambda)
-\Lambda_0(k^2)k^2,
\end{eqnarray}
where the function $\Lambda_0(k^2)$ defines the physical scales of the
system and characterizes the interaction.  The function  $\Lambda_0(k^2)$
  has to be chosen appropriately.

If we introduce Eqs. (\ref{55})  and (\ref{56}) into Eq. (\ref{a10}),  the
$\Lambda \to \infty$ limit can be taken immediately and one
obtains the exact renormalized result
\begin{eqnarray}
{K_R(k^2)} =- 1/\Lambda_0(k^2).\label{ab}
\end{eqnarray}
The physical scale(s) in $\Lambda_0(k^2)$ are to be identified with a
physical observable(s). If the problem is to be characterized by a single
physical scale, for example the scattering length $a$, it is appropriate
to take $\Lambda_0(k^2)$ to be independent of $k^2$: 
\begin{eqnarray}
\Lambda_0(k^2)
=-1/a. \nonumber
\end{eqnarray}
If it is to be characterized by two physical scales, such as the
scattering length $a$ and effective range $r_0$, it is natural to take 
\begin{eqnarray}
\Lambda_0(k^2)= -\frac{1}{a} + \frac{1}{2}r_0 k^2+ ... \nonumber
\end{eqnarray}
A third and more scales can be accomodated in a similar fashion. Hence at
low
energies one can accomodate  the full effective range expansion. The
general
solution for the renormalized $K$ matrix or its inverse at low energies is
a polynomial
in $k^2$.

Next we perform  dimensional regularization. From Eq.
(\ref{40}),  we find that the dimensional regularization of integrals
$I_L(k^2)$ for $d=3$  are all zero.  Here we use the energy-dependent bare
couplings 
\begin{eqnarray}\label{55x}
\frac{\lambda_1}{\lambda_2^2}&=& 
+\Lambda_0(k^2)k^4,\\
\label{56x}
\frac{1}{\lambda_2}&=& 
-\Lambda_0(k^2)k^2,
\end{eqnarray}
in Eq. (\ref{a10}), where the the functions $\Lambda_0(k^2)$ again  
define the
physical scales of the
system and characterizes the interaction. Then we obtain 
the finite
renormalized on-shell $K$ matrix given by Eq. (\ref{ab}). 
This result is identical with
the result obtained with cut-off regularization above. With the
flexibility introduced through the use of energy-dependent bare couplings,
the problem and inconsistency mentioned in Sec. I have been avoided.
Essentially, we have obtained Eq. (\ref{ab}) just by employing appropriate
bare
couplings and by introduzing subtractions in the divergent integrals
without choosing  a regularization scheme. In cut-off regularization the 
integrals $I_L(k^2, \Lambda) $ are energy-dependent divergent 
quantities, in dimensional regularization they are zero. This is the
only explicit difference between the bare couplings Eqs. (\ref{55}) and
(\ref{56}), 
and Eqs. (\ref{55x}) and (\ref{56x}), respectively. 
As we never have
to 
choose a specific regularization scheme, the results are identical.

\section{ Summary}

We have renormalized the $K$ matrices with potential (\ref{3}) by cut-off
and dimensional regularizations.  The solution of the dynamical problem in
these cases involves ultraviolet divergences. For this potential all
dimensionally regularized divergent integrals over Green functions are
identically zero. However, if the divergent integrals are appropriately
subtracted with the use of 
 energy-dependent bare couplings without choosing a specific
regularization procedure, we obtain  a finite result 
and 
both cut-off and dimensional regularization schemes lead to
equivalent renormalized results.  This suggests that once energy-dependent
bare couplings are chosen appropriately, the full nucleon-nucleon
potential
such as (\ref{div}) can be successfully renormalized using dimensional and
cut-off regularization schemes with both schemes leading to the same
renormalized result at low energies. However, it may not be easy to
perform an analogous renormalization for the full potential (\ref{div}) 
analytically. The difficulties that will arise in this task are expected
to be only technical in nature, and not questions of principle as
commented
in other investigations \cite{coh1,coh}.  The present study demonstrate
that renormalization with conventional cut-off and
dimensional regularizations in momentum space 
are
efficient tools for treating
divergent potentials in nonrelativistic quantum mechanics provided that  
appropriate energy-dependent bare couplings are employed.

We thank 
 the Conselho Nacional de Desenvolvimento Cient\'{\i}fico e Tecnol\'ogico, 
Funda\c c\~ao de Amparo \`a Pesquisa do Estado de S\~ao Paulo,  
Financiadora de Estudos e Projetos
of Brazil.


\begin{references}

\bibitem {wei}  S. Weinberg,  { Nucl. Phys. B} {\bf 363},   3 (1991);
{Phys. Lett. B} {\bf 251},   288 (1990); {\bf 295}, 114 (1992),
{Physica A} {\bf 96},  327
(1979).

\bibitem{eft3}  C. Ordon\'ez, L. Ray, and U. van Kolck, Phys. Rev. C {\bf
53}, 2086 (1996).


\bibitem{coh1} D. R. Phillips, S. R. Beane, and T. D. Cohen, Nucl Phys.
{\bf A632}, 445 (1998); Ann. Phys. (N.Y.) {\bf 263}, 255 (1998).


\bibitem{bir}K. G. Richardson, M. C. Birse, and J. A. McGovern, preprint,
hep-ph/9708435; M. C. Birse, J. A. McGovern, and K. G. Richardson, preprint,
hep-ph/9807302. 

\bibitem{eft2} D. B. Kaplan,   M J. Savage,   and  M. B. Wise,   { Nucl.
Phys.  B} {\bf 478},   629 (1996);
D. B. Kaplan, {Nucl.
Phys. B} {\bf 494}, 471  (1997).


\bibitem{ad} S. K. Adhikari and A. Ghosh,    { J. Phys. A} {\bf 30},  6553
(1997).

\bibitem{adx} S. K. Adhikari and  T. Frederico,  { Phys. Rev.
Lett.} {\bf 74},  4572 (1995);  S. K.  Adhikari,  T. Frederico,  and I. D.
Goldman,  {\it ibid.} {\bf 74},   487 (1995); C. F.  de Araujo,  Jr.,  L.
Tomio,  S. K. Adhikari,   and  T.  Frederico,   {J. Phys. A} {\bf 30},  4687
(1997).



\bibitem{lat} S. K. Adhikari,  T. Frederico and R. M. Marinho,
{J. Phys. A} {\bf 29},  7157 (1996).

\bibitem{sad}A. Ghosh, S. K. Adhikari, and B. Talukdar, Phys. Rev. C
{\bf 58}, 1913 (1998). 

\bibitem {jac} I. Mitra, A. DasGupta, and B. Dutta-Roy,
Am. J. Phys. {\bf 66}, 1101 (1998); 
C.  Manuel and R. Tarrach, { Phys. Lett. B} {\bf 328},   113 (1994); 
T. J.  Fields,  K.
S. Gupta,   and J. P. Vary,   {Mod. Phys. Lett. A} {\bf 11}, 2233 1996;  
G.
Amelino-Camelia,   { Phys. Lett. B} {\bf326},  282 (1994); {Phys. Rev.  D}
{\bf51},  2000 (1995); 
H. El Hattab and J. Polonyi, Ann. Phys. (N.Y.) {\bf 268}, 207 (1998).
M. Luke and A. V. Manohar, Phys. Rev. D {\bf 55}, 4129 (1997).


\bibitem{kap}D. B. Kaplan,   M J. Savage,   and  M. B. Wise,   { Nucl.
Phys.  B} {\bf 534},   329 (1998); Phys. Lett. B 
{\bf 424}, 390 (1998).

\bibitem{coh}T. D. Cohen and J. M. Hansen, Phys. Lett. B
{\bf 440}, 233 (1998).


\bibitem{new}  J. Gegelia,
{Phys. Lett. B}
{\bf 429}, 227 (1998); D. R. Phillips, S. R. Beane, and M. C. Birse, 
preprint, hep-th/9810049. 


















\end{references}
\end{document}